\documentclass[twocolumn,showpacs,superscriptaddress,amsmath,amssymb]{revtex4}
\usepackage{amsmath}
\usepackage{verbatim}
\usepackage{psfrag}
\usepackage{color}
\usepackage{stmaryrd}
\usepackage{mathrsfs}
\usepackage{graphicx}
\usepackage{dcolumn}
\usepackage{bm}
 
\def\bd{\begin{document}} \def\ed{\end{document}}
\def\bmp{\begin{minipage}} \def\emp{\end{minipage}}
\def\bcc{\begin{center}} \def\ecc{\end{center}}     \def\npg{\newpage}
\def\beq{\begin{equation}} \def\eeq{\end{equation}} \def\hph{\hphantom}
\def\be{\begin{equation}} \def\ee{\end{equation}} \def\r#1{$^{[#1]}$}
\def\n{\noindent} \def\ni{\noindent} \def\pa{\parindent}
\def\hs{\hskip} \def\vs{\vskip} \def\hf{\hfill} \def\ej{\vfill\eject}
\def\cl{\centerline} \def\ob{\obeylines}  \def\ls{\leftskip}
\def\underbar#1{$\setbox0=\hbox{#1} \dp0=1.5pt \mathsurround=0pt
   \underline{\box0}$}   \def\ub{\underbar}    \def\ul{\underline}
\def\f{\left} \def\g{\right} \def\e{{\rm e}} \def\o{\over} \def\d{{\rm d}}
\def\vf{\varphi} \def\pl{\partial} \def\cov{{\rm cov}} \def\ch{{\rm ch}}
\def\la{\langle} \def\ra{\rangle} \def\EE{e$^+$e$^-$} \def\pt{p_{\rm t}}
\def\dt{\delta}   \def\ie{{\it i.e.\;}}   \def\cf{{\it cf.\;}}
\def\bitz{\begin{itemize}} \def\eitz{\end{itemize}}
\def\btbl{\begin{tabular}} \def\etbl{\end{tabular}}
\def\btbb{\begin{tabbing}} \def\etbb{\end{tabbing}}
\def\beqar{\begin{eqnarray}} \def\eeqar{\end{eqnarray}}
\def\\{\hfill\break} \def\dit{\item{-}} \def\i{\item}
\def\bbb{\begin{thebibliography}{9} \itemsep=-1mm}
\def\ebb{\end{thebibliography}} \def\bb{\bibitem}
\def\bpic{\begin{picture}(260,240)} \def\epic{\end{picture}}
\def\akgt{\noindent{Acknowledgements}}
\def\fgn{\noindent{\bf\large\bf figure captions}}
\def\lan{\langle}
\def\ran{\rangle}
\def\p{\pi}
\def\ifmath#1{\relax\ifmmode #1\else $#1$\fi}%
\def\rc{\ifmath{{\mathrm{c}}}}
\def\cut{\ifmath{{\mathrm{cut}}}}
\def\rF{\ifmath{{\mathrm{F}}}}
\def\rK{\ifmath{{\mathrm{K}}}}
\def\rp{\ifmath{{\mathrm{p}}}}
\def\rt{\ifmath{{\mathrm{t}}}}
\def\LAB{\ifmath{{\mathrm{LAB}}}}
\def\cut{\ifmath{{\mathrm{cut}}}}
\def\beq{\begin{equation}}
\def\eeq{\end{equation}}
\def\us{^{(s)}}  \def\bea{\begin{eqnarray}} \def\eea{\end{eqnarray}}
\def\nbr{\nonumber} \def\e{{\rm e}} \def\dt{\delta} \def\D{\Delta}
\def\r{\rho}  \def\unln{\underline}
\newcommand{\cinst}[2]{$^{\mathrm{#1}}$~#2\par}
\newcommand{\crefi}[1]{$^{\mathrm{#1}}$}
\newcommand{\crefii}[2]{$^{\mathrm{#1,#2}}$}
\newcommand{\crefiii}[3]{$^{\mathrm{#1,#2,#3}}$}
\newcommand{\HRule}{\rule{0.5\linewidth}{0.5mm}}
\def\yw{Y_{\rm w}}  \def\ew{\eta_{\rm w}}  \def\de{\delta\eta} \def\Bs{B_{\rm s}}
\def\bs{\boldsymbol}  \def\lm{\lambda}     \def\tc{T_{\rm cr}}
\def\fc{\frac}   \def\d{{\rm d}}

\begin{document}

\title{Critical phenomena in disc-percolation model
and its application to\\ relativistic heavy ion collisions}

\author{Ke Hongwei}

\affiliation{Institute of Particle Physics, Huazhong Normal
University, Wuhan 430079, China}

\author{Xu Mingmei} \affiliation{Institute of
Particle Physics, Huazhong Normal University, Wuhan 430079, China}
\affiliation{Key  Lab of Quark and Lepton Physics (Huazhong Normal
University), Ministry of Education, China}

\author{Liu Lianshou}
\email{liuls@iopp.ccnu.edu.cn} \affiliation{Institute of Particle
Physics, Huazhong Normal University, Wuhan 430079, China}
\affiliation{Key Lab of Quark and Lepton Physics (Huazhong Normal
University), Ministry of Education, China}

\def\ssnn{\sqrt {s_{NN}}}

\begin{abstract}
Through studying the critical phenomena in continuum-percolation of
discs, we find a new approach to locate the critical point, i.e.
using the inflection point of $P_\infty$ as an evaluation of the
percolation threshold. The susceptibility, defined as the derivative
of $P_\infty$, possess finite-size scaling property, where the
scaling exponent is the reciprocal of $\nu$ --- the critical
exponent of correlation length. The possible application of this
approach to the study of the critical phenomena in relativistic
heavy ion collisions is discussed. The critical point for
deconfinement can be extracted by the inflection point of $P_{\rm
QGP}$ --- the probability for the event with QGP formation. The
finite-size scaling of its derivative can give the critical exponent
$\nu$, which is a rare case that can provide an experimental measure
of a critical exponent in heavy ion collisions.
\end{abstract}


\maketitle

\section{Introduction}
\label{sec1}

The general feature of the phase diagram of strongly interacting
matter has become increasingly well
established~\cite{lattice-review}. Following a region of crossover
around a temperature of 170 - 190 MeV at zero baryon chemical
potential $\mu_B$, increasing $\mu_B$ leads to a critical point,
beyond which the system shows a first order transition from confined
to deconfined phase.

Recently, the interest in QCD critical point progressively rises.
With the plan of RHIC low energy scan and GSI new facility, the
study of relativistic heavy ion collisions becomes concentrating on
searching for the critical point and observing the relevant critical
phenomena, in particular measuring the critical exponents.

On the theoretical side, studying the critical phenomena in chiral
symmetry restoration is available~\cite{critical-chiral}. Many
discussions about universality have been active~\cite{universality}.
The chiral condensate and Polyakov loop are proposed as order
parameters for chiral restoration and color deconfinement,
respectively. However, it is a pity that both of these variables can
not be directly measured by experiment. What observable should we
measure, how to locate the critical point and extract the
corresponding critical exponents in heavy ion collisions have not
been clear so far.

On the experimental side, what we observed in heavy ion collisions
is the deconfined partonic degree of freedom~\cite{exp-deconfine}.
Despite the fact that confinement is a long standing problem not
solved by theory, studying the critical phenomena in deconfinement
phase transition is more realistic for experiments. Since chiral
symmetry is hard to measure and deconfinement is easier to observe,
we propose to phenomenologically study the critical phenomena from
the deconfinement aspect and give some hints for experiments.

Principally speaking, it is impossible to get the critical point
because of the limited system size in relativistic heavy ion
collisions. In this paper, by means of studying the critical
phenomena in finite-size continuum-percolation of disks, we find
that $P_{\infty}$, the probability for event with infinite cluster,
has an inflection point, which is a good approximation for the
critical point. Finite-size scaling method is further used to
extract the critical exponents $\nu$ from the distribution of
susceptibility. Since percolation has some resemblance with
deconfinement~\cite{Satz-deconfinement}, a possible application of
the method used in percolation to the study of the critical
phenomena in relativistic heavy ion collisions is discussed.

\section{The continuum-percolation of disc}
\label{sec2}

The continuum-percolation problem has various formulations, among
which the {\it problem of spheres} is most popular. Equally sized
spheres are placed at random in a substrate. The spheres support the
transport and the substrate does not. When we put enough spheres in
the substrate, the overlapping spheres form an infinite cluster, and
the system is able to support a long-range current. This model has
been well studied and was used to describe hopping conduction in
doped semiconductors~\cite{Shk-1980} and phase transitions in
ferromagnetics~\cite{Abr-1980}. Such a model together with its
critical behavior also bare some resemblance with the transition
from hadron gas to quark-gluon plasma.

In QCD, hadrons are color-singlet bound states of more basic colored
objects --- quarks and gluons. Hadronic matter, consisting of
colorless constituents of hadronic dimensions, can turn at high
temperature and/or density to a quark-gluon plasma of colored quarks
and gluons as constituents in a much larger volume. This
deconfinement transition leads to a color-conducting state and thus
is the QCD counterpart of the insulator-conductor transition in
atomic matter. Suppose hadrons have an intrinsic size. At low
density, we have a hadron gas. When this becomes so dense that the
formation of an infinite cluster occurs, it turns to a quark-gluon
plasma. In this case, the connectivity (cluster formation)
determines the different states of many-body systems. The lesson
learned from spin systems also indicates that cluster formation and
associated critical behavior are the more general
feature~\cite{Satz-deconfinement}.

\subsection{Model parameters}
In relativistic heavy ion collisions the incoming nuclei become
discs of vanishing thickness due to Lorentz contraction and the
collision region is two-dimensional. So we turn to study
two-dimensional percolation.

In a two-dimensional system, spheres become discs. Discs with radius
$a$, called {\it cells}, distribute randomly in the system of a big
disc with radius $R$. The ratio $R/a$ determines the {\it system
size} and is denoted by $s$. The {\it control variable} for
percolation is $\eta$ --- the ratio of the total area of all the
cells and the area of the big disc, i.e.
\begin{equation}\label{def_eta}
    \eta = \frac{N \pi a^2}{\pi R^2} = \frac{N a^2}{R^2},
\end{equation}
where $N$ is the number of cells. The increase of cell number $N$
leads to the increase of $\eta$. The value of $\eta$ at which an
infinite cluster appears is the critical point and is denoted by
$\eta_c$.

The definition for infinite cluster varies. In finite system a
cluster spanning the system is called an infinite cluster. Let
$(r_i, \theta_i)$ be the polar coordinates of cell $i$. If $r_{i}
\in [R-a, R]$ we call the cell $i$ as a {\it boundary cell}. If
cells $i$ and $j$ are two boundary cells belonging to the same
cluster, and $\left|\theta_i - \theta_j \right| \geq \theta_0$ then
we say the current cluster is infinite. Here $\theta_0$ is a
parameter.

Thus the control variable is $\eta$ and the model parameters are $s$
and $\theta_0$.

\subsection{Locating the critical point}
\begin{figure}
\begin{center}
\includegraphics[scale=0.44]{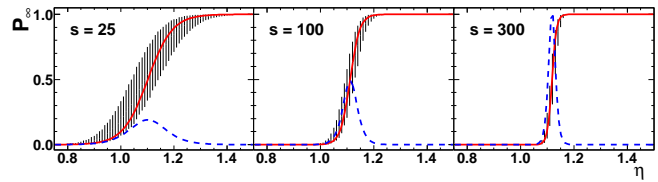}
\caption{(Color online) $P_\infty$ for different system size. From
left to right the system sizes are 25, 100, 300, respectively. The
error bar shown is systematic error induced by the variation of
parameter $\theta_0$. Statistical errors are small and not shown.
The dashed curves are the susceptibility defined in Eq.\;(4)}
\label{fig_pinf}
\end{center}
\end{figure}

When $s$ and $\theta_0$ take certain values respectively, {\it the
probability for the event with infinite cluster}, denoted by
$P_\infty$, is an increasing function of $\eta$. For infinite system
size, $P_\infty = 0$ holds for all $\eta < \eta_c$. When $\eta \geq
\eta_c$, $P_\infty$ picks up and sharply increases to unity. System
of infinite size is unrealistic. $P_\infty$ for finite system of
three different sizes are measured and shown in Fig.~\ref{fig_pinf}.

Different parameters, e.g. $\theta_0 = 90^\circ, 135^\circ \text{
and } 175^\circ $, are tried and the induced systematic errors are
shown as error-bars in the figures. The solid curves in
Fig.~\ref{fig_pinf} are the fitting result by using the function
\begin{equation}\label{fitfunc_pinf}
    P_\infty(\eta) = \frac{1 + \tanh\left[ c_1\left( \eta - c_2 \right)
    \right]}{2}.
\end{equation}
It can be seen that for all three cases the fitting function always
lies within the systematic errors. The fitting function, i.e.
Eq.(\ref{fitfunc_pinf}), has an inflection point at $\eta = c_2$. As
$s$ increases, $P_\infty$ tends to a step function and the
inflection point $c_2$ tends to $\eta_c$. For finite system size,
the inflection point $\eta = c_2$ can be used as an evaluation of
$\eta_c$.

\begin{figure}
\begin{center}
\includegraphics[scale=0.3]{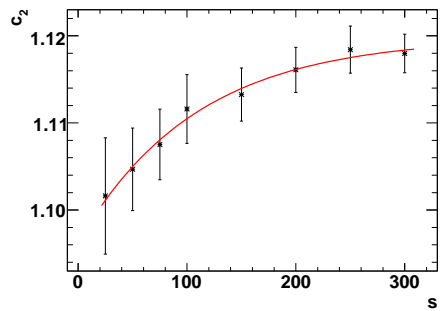}
\caption{(Color online) The behavior of $c_2$ versus system size. A
tendency to saturation can be seen.} \label{fig_beta}
\end{center}
\end{figure}

The behavior of $c_2$ versus different system size is shown in
Fig.~\ref{fig_beta}. A tendency to saturation can be seen. The
saturation value of $c_2$ is $1.1198 \pm 0.0047$, i.e. the
percolation threshold is $\eta_c = 1.1198 \pm 0.0047$.

For a nucleus-nucleus collision, the contraction induced by
relativistic motion makes the colliding nucleus like a disc, in
which disc-like nucleons percolate. For central Au-Au collision, the
radius of the interacting region is about 7 fm and the hard core of
each nucleon is about 0.1 fm, resulting in a system size of about
70. The inflection point for $s=70$ is about 1.1054. Using it as an
approximation for $\eta_c$, the error is 1.3\%. So using the
inflection point as an approximation for critical point is easy to
measure and exact enough.

In heavy ion collisions, when hadrons connect to form infinite
cluster, color conducts in the system and quark gluon plasma is
formed. If the formation of QGP can be signaled in each collision,
then the probability for the event with infinite cluster, denoted by
$P_\infty$, becomes the probability for the collision with QGP
------ $P_{\rm QGP}$. $P_{\rm QGP}$ should be measurable in experiment,
provided that an available signal is found, and thus in principle
the critical point can be obtained according to the inflection point
of $P_{\rm QGP}$.

$P_\infty$ is also referred to by some authors~\cite{Satz-cumulant}
as {\it percolation cumulant}, in analogous to the Binder
cumulant~\cite{Binder-cumulant}, which will intersect at the
critical point for different system sizes. Extracting the
intersection point of $P_\infty$ is a method to estimate the
critical point. However, to use the inflection-point method proposed
above only needs the collisions of one kind of ion at various
energies, while to use the intersecting-point method need the
collision at different energies of more than one kind of ion.
Therefore, the inflection-point method is more realistic and the
value of $\eta_{\rm c}$ obtained for system sizes as small as Pb
($s=71$), Au ($s=70$) is a good estimation with error about 1.3\%.

\begin{figure}
\includegraphics[scale=0.3]{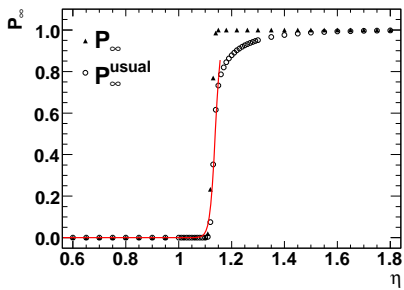}
\caption{(Color online) $P^{\mathrm{usual}}_{\infty}$ (open circles)
as a function of $\eta$ with system size $s=1000$. The solid curve
is the fit of the lower part to Eq.~(\ref{fitfunc_pinfu}). For
comparison the $P_{\infty}$ for $s=1000$ is also shown as triangles
. } \label{fig_D3}
\end{figure}

The inflection-point method is also applicable for the P-infinity
defined in another way.  In Ref.~\cite{book1-perco} P-infinity is
defined as \emph{the probability of a cell belonging to an infinite
cluster}, which will be denoted by $P^{\mathrm{usual}}_{\infty}$ in
the following. Fig.~\ref{fig_D3} shows $P^{\mathrm{usual}}_{\infty}$
as a function of $\eta$ when $s=1000$. In case of $s=1000$, the
system is a  big one and $P_\infty$ rises like a step function, cf.
Fig.~\ref{fig_D3}, but $P^{\mathrm{usual}}_{\infty}$ has a step
function sharp rise only at the bottom side ($P_\infty^{\rm
usual}=0$), while at the upper side $P_\infty^{\rm usual}$ tends to
unity smoothly.

In this case, we could only fit the left part of $P_\infty^{\rm
usual}(\eta)$ to Eq.~(\ref{fitfunc_pinf}), or alternatively, to a
modified fitting function
\begin{equation}\label{fitfunc_pinfu}
     P_\infty^{\rm usual}(\eta) =
     c_1 \left[ c_2 + \tanh\left( c_3 \left( \eta-c_4 \right) \right)
    \right].
\end{equation}
The fitting is shown as solid curve in Fig.~\ref{fig_D3}. The
inflection point of this half-fitted curve can also be used as an
approximation of the critical point.

\subsection{Susceptibility and the critical exponent $\nu$ }

Susceptibility is defined by the response of the system to small
external forces. We define the susceptibility in percolation as the
derivative of $P_\infty$ with respect to the control variable
$\eta$, i.e.
\begin{equation}\label{def_kappa}
    \kappa(\eta, s) = \frac{\partial P_\infty(\eta, s)}{\partial
    \eta}.
\end{equation}
$\kappa$ shows a peak at the inflection point of $P_\infty$, see the
dashed line in Fig.~\ref{fig_pinf}. Since $P_\infty$ becomes a step
function when $s$ tends to infinity, the peak of $\kappa$ will
become higher and narrower and finally diverge at $\eta_c$.

In realistic case, the system is of finite size determined by the
size of the colliding nuclei. The finite-size scaling
method~\cite{finite-size-scaling}, investigating the scaling of
quantities at $\eta_c$ as a function of system size, is adopted to
extract values for the critical exponents.

The finite-size scaling of $P_\infty$ suggests~\cite{Stauffer}
\begin{equation}\label{P_scale}
P_\infty = \Phi\left[\left(\eta-\eta_c\right)s^{1/\nu}\right]\quad
\text{for large } s, \quad \eta \rightarrow
    \eta_c.
\end{equation}
Denoting $X=(\eta-\eta_c)s^{1/\nu}$, $\Phi(X)$ is some function of
$X$. $\nu$ is the critical exponent of correlation length $\xi$,
i.e.
\begin{equation}\label{def_nu}
    \xi \propto \left| \eta - \eta_c \right|^{-\nu} \quad  \text{for } \eta \rightarrow
    \eta_c.
\end{equation}
Then, the susceptibility of $P_\infty$  is
\begin{equation}
\kappa(\eta, s) = \frac{\partial P_\infty(\eta, s)}{\partial
    \eta}=s^{1/\nu}\frac{{\rm d}\Phi}{{\rm d} X}.
\end{equation}
When $\eta=\eta_c$, $\frac{{\rm d}\Phi}{{\rm d} X}|_{X=0}$ is a
constant and
\begin{equation}\label{kappa_scale}
\kappa(\eta_c, s)\propto s^{1/\nu}\quad  \text{for large } s.
\end{equation}
So the divergence behavior of $\kappa$ near $\eta_c$ is related with
the critical exponent of correlation length.

\begin{figure}
\includegraphics[scale=0.3]{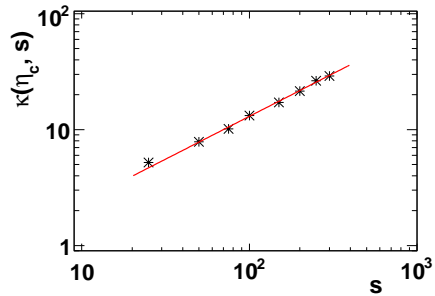}
\caption{(Color online) The distribution of susceptibility $\kappa$
at critical point $\eta_c$ as a function of system size $s$. The
solid line is a fitting by
Eq.~(\ref{kappa_scale}).}\label{fig_kappa_fs}
\end{figure}

Using the critical point $\eta_c$ extracted from $P_\infty$, we
evaluate $\kappa(\eta_c, s)$ as a function of $s$, shown in
Fig.~\ref{fig_kappa_fs}. Fitting it to Eq.~(\ref{kappa_scale}) we
obtain $1 / \nu = 0.739 \pm 0.041$, $\nu=1.353\pm0.075$. The
exponent $\nu$ obtained is within 1.5\% difference with the result
from other calculations for two-dimension
percolation~\cite{book1-perco}.

The critical exponent of correlation length plays a special role in
the theory of critical phenomena, because the scaling behavior of
other quantities depends on the relative magnitude of correlation
length and system size. Hence, the critical exponent of correlation
length appears in the scaling relation of various quantities. By
definition, correlation length is the distance at which correlation
function reduces to $1/{\rm e}$~\cite{book1-perco}. However,
correlation function usually has non-monotonic behavior, e.g. in
case of liquid, correlation function shows damping
oscillation~\cite{book-liquid}, which makes the correlation length
hard to be measured. Now we see that using the scaling behavior of
susceptibility the critical exponent of correlation length can
easily be obtained~\cite{Stauffer}.

\section{Application to relativistic heavy ion collision}

\begin{figure}
\includegraphics[scale=0.3]{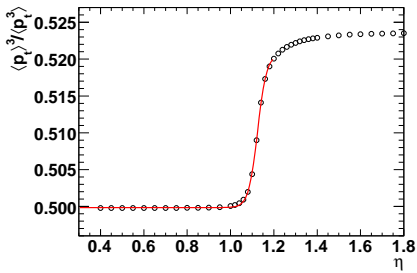}
\caption{(Color online) $D_3$ as a function of $\eta$ with system
size $s=500$. Red line is a fitting by Eq.(\ref{fitfunc_pinfu}).}
\label{fig_D3_s500}
\end{figure}

Recently, a kind of scaled third order moment of transverse
momentum~\cite{dujx-moment}
\begin{equation}\label{def_d3}
    D_3 = \frac{\left\langle p_t \right\rangle^3}{\left\langle p_t^3 \right\rangle}
\end{equation}
has been proposed as a possible signal of critical-point (CP) of QCD
phase diagram. By imposing a temperature gradient to the two
dimensional continuum percolation of discs and assuming the
transverse momentum for each cell takes the thermal momentum
determined by the temperature, $D_3(\eta)$ behaves like a step
function, similar to that of $P_\infty^{\rm usual}(\eta)$ described
above, cf. Fig.~\ref{fig_D3_s500} and Fig.~\ref{fig_D3}. This
similarity makes the methods developed in Sec.\ref{sec2} applicable
to $D_3$, i.e. we can locate the CP and determine the critical
exponent of a susceptibility related to $D_3$.

\begin{figure}
\includegraphics[scale=0.3]{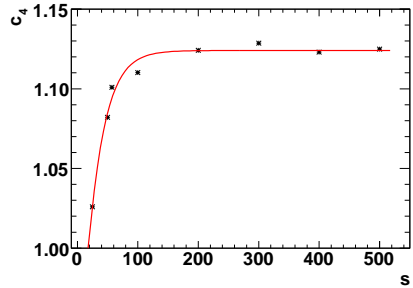}
\caption{(Color online) The behavior of $c_4$ versus different
system size. A saturation is seen.} \label{fit_etac_D3}
\end{figure}

A typical result for $D_3$ with a half-fit to
Eq.~(\ref{fitfunc_pinfu}) is shown in Fig.~\ref{fig_D3_s500}. The
inflection point $c_4$ is regarded as the approximation of critical
point $\eta_c$. With the increasing of system size, the inflection
point $c_4$ has an asymptotic behavior which is shown in
Fig.~\ref{fit_etac_D3}. The saturation value indicates
$\eta_c=1.124$.

\begin{figure}
\begin{center}
\includegraphics[scale=0.3]{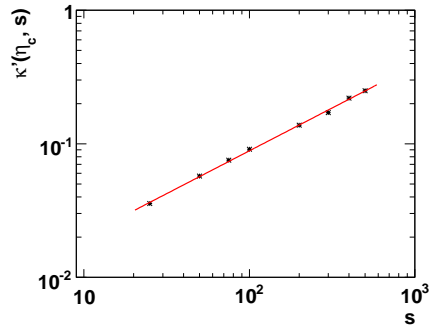}
\caption{(Color online) The distribution of susceptibility $\kappa'$
at critical point $\eta_c$ as a function of system size $s$. The
solid line is a fitting to Eq.~(\ref{def_nu_prime}).}
\label{fig_kappa_prime_fs}
\end{center}
\end{figure}

Similar to Eq.(\ref{def_kappa}) we define the susceptibility related
to $D_3$ as
\begin{equation}\label{def_kappa_prime}
  \kappa'(\eta, s) = \frac{\partial D_3(\eta, s)}{\partial \eta}, \\
\end{equation}
and a critical exponent $\nu'$ for this thermal system as
\begin{equation}\label{def_nu_prime}
   \kappa'(\eta_c,s) \propto s^{1/\nu'}.
\end{equation}
Investigating the finite-size scaling of $\kappa'$, $1/ \nu' =
0.642$ is obtained, c.f. Fig.~\ref{fig_kappa_prime_fs}.

\section{Conclusion}
We study the critical phenomena in continuum percolation of discs.
It is shown that using the inflection point of $P_\infty$ or
$P_\infty^{\rm usual}$ as an estimation for percolation threshold is
a good approximation. The critical exponent $\nu$ is extracted from
the distribution of susceptibility by using finite-size scaling
method.

For heavy ion collisions, different colliding nuclei have been
tried, e.g. Pb, Au, Cu, S, C etc., which play the role of system
size in percolation. The colliding energy plays the role of control
variable. $P_\infty$ in percolation corresponds to the probability
of events with QGP in heavy ion collision experiment, denoted as
$P_{\rm QGP}$. Studying the inflection point of $P_{\rm QGP}$ as a
function of $\sqrt{s_{NN}}$, the critical point can be extracted. A
susceptibility defined as the derivative of $P_{\rm QGP}$ determines
the critical exponent $\nu$ which is a rare case that can
experimentally measure the critical exponent in heavy ion
collisions.

The analogue between percolation and deconfinement is reasonable.
The approach studied here is worth trying in the future study of the
critical phenomena in real relativistic heavy ion collisions.

{\bf Acknowledgement} \ This work is supported by NSFC under
projects No.10775056, 10835005 and 10847131.

\end{document}